\documentclass{IEEEtran}
\usepackage{cite}
\usepackage{amsmath,amssymb,amsfonts}
\usepackage{graphicx}
\usepackage{textcomp,nicefrac}
\usepackage{siunitx} 
\def\BibTeX{{\rm B\kern-.05em{\sc i\kern-.025em b}\kern-.08em
T\kern-.1667em\lower.7ex\hbox{E}\kern-.125emX}}
\begin{document}
\title{Stable Charge Collection and Sub-45~ps Time Resolution in a 4H-SiC PIN Detector Irradiated With Low Fluence 16.5 MeV/u Ta Ions}

\author{Jingxuan. He, Congcong. Wang,  Yi. Zhan, Zhenyu. Jiang, Xiyuan. Zhang and Xin. Shi
\thanks{This work is supported by the National Natural Science Foundation of China (Nos. 12305207, 12375184 and 12405219) and support from CERN DRD3 Collaboration. (Corresponding author: Congcong Wang)}
\thanks{Jingxuan He is with the Institute of High Energy Physics, Chinese Academy of Sciences, Beijing 100049, China, and also with Qufu Normal University, Rizhao, 276826, China.}
\thanks{Congcong Wang is with the Institute of High Energy Physics, Chinese Academy of Sciences, Beijing 100049, China, and also with the State Key Laboratory of Particle Detection and Electronics, Beijing 100049, China. (e-mail: wangcc@ihep.ac.cn)}
\thanks{Yi Zhan is with Qufu Normal University, Rizhao, 276826, China.}
\thanks{Zhenyu Jiang is with the Institute of High Energy Physics, Chinese Academy of Sciences, Beijing 100049, China, and also with Liaoning University, Shenyang 110136, China.}
\thanks{Xiyuan Zhang and Xin Shi are with the Institute of High Energy Physics, Chinese Academy of Sciences, Beijing 100049, China.}}
\maketitle

\maketitle

\begin{abstract}
A silicon carbide PIN detector was fabricated and its radiation tolerance under Ta heavy ion irradiation of 2370 MeV was evaluated. Its electrical properties, charge collection performance and time resolution of  $\beta$-particles ($^{90}$Sr) are reported. The leakage currents for unirradiated and irradiated 4H-SiC PIN detectors are $1.47 \times 10^{-10}$~A @ 300 V and 1.49~$\times$ 10$^{-10}$A@ 300 V.
The effective doping concentrations for unirradiated and irradiated 4H-SiC PIN detectors are $6.23\times 10^{13}$~cm$^{-3}$ and $6.13\times 10^{13}$~cm$^{-3}$. The irradiated detector exhibits good electrical performance and stable device architecture. The 4H-SiC PIN detector exhibits a charge collection efficiency (CCE) of 99.24\% under Ta Heavy Ion Irradiation. The time resolutions of the detector before and after irradiation are 40 ps and 45 ps, respectively. Experimental results indicate that the CCE and time resolution performance exhibit good stability before and after irradiation. These results demonstrate stable performance under Ta heavy ion irradiation, highlighting the detectors potential for radiation-hard applications in high-energy physics, space missions, and nuclear reactor monitoring.
\end{abstract}


\section{Introduction}
\label{sec:introduction}
\IEEEPARstart{I}{n} fields such as nuclear physics experiments, space exploration, Medical physics, nuclear safety, particle physics and accelerators and advanced reactor monitoring, detectors are inevitably exposed to intense radiation fields composed of heavy ions for extended periods\cite{mao2024}. High-energy physics collider experiments require particle track reconstruction and time-of-flight measurements, demanding fast time response and stable charge collection to be maintained after heavy-ion irradiation. Space heavy-ion detection requires the ability to maintain stable performance in energy spectrum measurements of high-Z particles (such as Fe and Ta) in galactic cosmic rays during long-term space missions after heavy-ion irradiation. Solar energetic particle event monitoring for heavy-ion bursts requires fast response and high reliability under heavy-ion irradiation\cite{Cao2025}. Particle flux measurements in fusion device diagnostics (e.g., ITER) require detectors to withstand heavy-ion irradiation. Monitoring of actinides in nuclear waste requires detectors to maintain long-term stability in detecting radionuclide activity under heavy-ion irradiation. Conventional silicon semiconductor detectors, limited by their narrow bandgap and low displacement damage threshold\cite{Cheng2025,Moll2018,Messenger1992}, suffer from dramatically increased dark current and severely degraded energy resolution under extreme-fluence irradiation, making it increasingly difficult to meet the ever-growing demands for radiation tolerance. The operational reliability of particle detectors under heavy-ion irradiation has emerged as one of the critical bottlenecks limiting further technological breakthroughs. Heavy ions, characterized by high linear energy transfer and strong displacement damage capability, impose extremely stringent demands on detector materials and device structures. Silicon carbide (SiC), as a wide-bandgap semiconductor material, possesses theoretically superior radiation resistance to silicon devices due to its high atomic displacement energy, high thermal conductivity, and excellent carrier saturation drift velocity\cite{DeNapoli2022,Chabi2020,Liu2015}. Silicon carbide detectors are widely recognized as one of the most promising radiation-hardened semiconductor detectors due to their superior intrinsic physical properties.

Time resolution is a core metric of a particle detector's ability to respond to transient or rapidly sequential events, directly determining its applicability and data quality in high-count-rate scenarios. The high saturation electron drift velocity and wide bandgap characteristics of silicon carbide enable faster carrier transport and a thinner depletion layer, which determine its fast time response capability. In addition, investigating the synergistic degradation patterns of charge collection efficiency and temporal resolution in silicon carbide detectors under heavy-ion irradiation, and revealing their intrinsic correlation with irradiation-induced defects, constitute a scientific prerequisite for evaluating device radiation tolerance and guiding performance optimization. However, systematic experimental data and in-depth theoretical understanding are still lacking regarding the degradation of electrical performance and charge collection efficiency (CCE), as well as the damage evolution of temporal resolution characteristics, in SiC detectors under heavy-ion irradiation (e.g., particles with high linear energy transfer such as Ta). Systematically evaluating the effects of heavy-ion irradiation on the electrical performance, charge collection efficiency, and temporal resolution of silicon carbide detectors is of crucial scientific value and practical significance for advancing the engineering application of such detectors in extreme radiation environments\cite{Tudisco2025,DeNapoli2022}.

The 4H-SiC PIN detector presented in this work was designed and fabricated by Congcong Wang(Institute of High Energy Physics, Chinese Academy of Sciences). In this work, the electrical characteristics were systematically evaluated in current-voltage (I-V) and one over (capacitance C) squared-voltage ($1/C^{2}\text{-}V$) curves of the 4H-SiC PIN detector before and after 16 MeV/u Ta ions at a total fluence with $1\times10^{9}~\mathrm{cm}^{-2}$. The differences in CCE and time resolution of its were compared and analyzed.

\section{Detector fabrication and irradiation conditions}

\subsection{Epitaxial structure and detector fabrication}

The SiC PIN detector employs a fully epitaxial vertical PIN architecture, the epitaxial structure  from bottom to top includes:

1) The conductive N-type 4H-SiC substrate with a thickness of 350~$\mu$m.

2) The lightly doped N-epi layer with a nitrogen ion doping concentration of 5 $\times$ 10$^{13}$ cm$^{-3}$ and a thickness of 50~$\mu$m.

3) The P++ layer with an aluminum ion doping concentration of 2 $\times$ 10$^{19}$ cm$^{-3}$ and a thickness of 0.6 $\mu$m.

\begin{figure}[ht]
    \centering
    \includegraphics[width=1\linewidth]{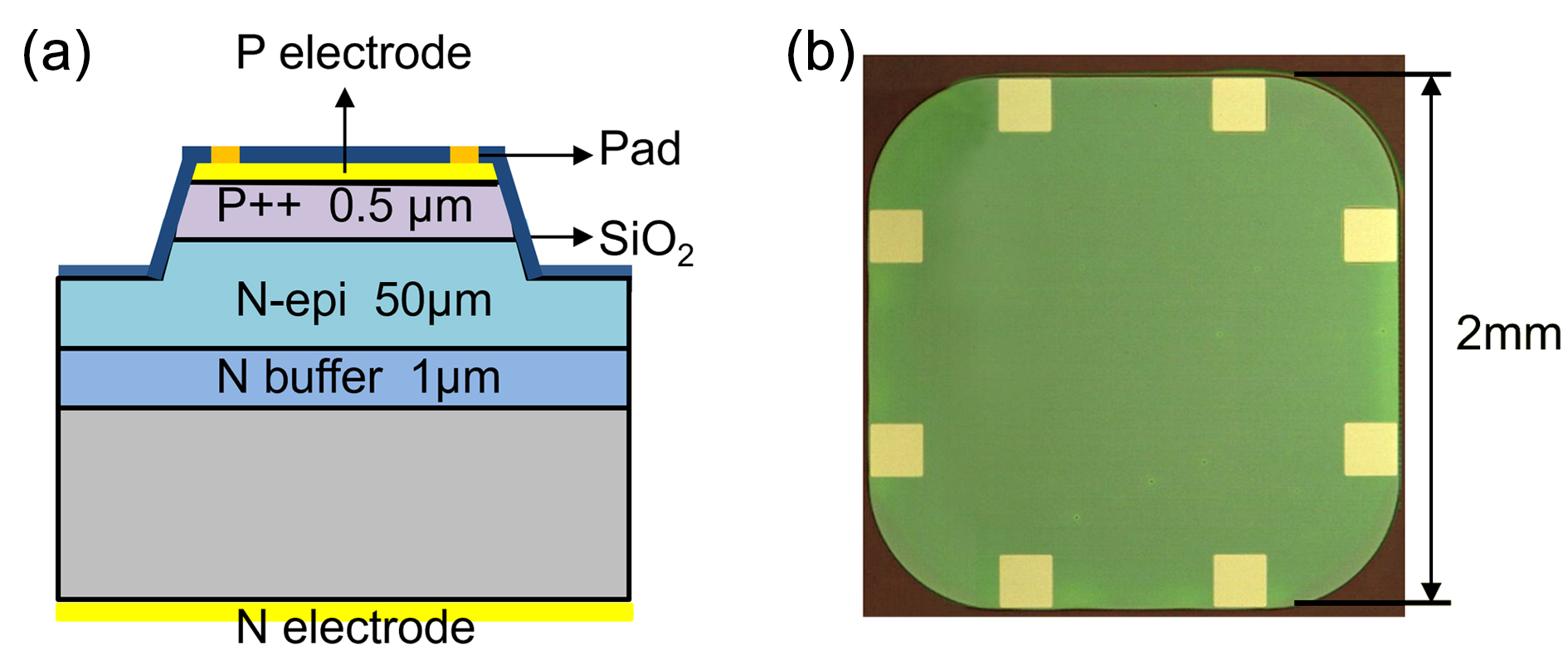}
    \caption{(a) Cross-sectional structure of the 4H-SiC PIN detector. (b) Real 4H-SiC PIN detector image.}
    \label{fig:detector}
    
\end{figure}

The structure of the 4H-SiC PIN detector includes P electrode, SiO$_{2}$ passivation layer, P++ layer, N-epi layer, N buffer layer, conductive N-type 4H-SiC substrate and N electrode as shown in Fig.~1. The size of the detectors is 2~mm $\times$ 2~mm. The fabrication process of the 4H-SiC PIN detector mainly includes lithography, etching, electron beam evaporation, magnetron sputtering and rapid thermal annealing. The etching depth of the epitaxial structure exceeds 0.6~$\mu$m to ensure complete removal of the P++ layer. A Ni/Ti/Al=50 nm/15 nm/80 nm electrode stack is deposited on both the P++ layer and the N-type substrate using electron beam evaporation. Rapid thermal annealing is 850~$^\circ$C for 2 minutes to form P-ohmic contacts. Then, a SiO$_{2}$ layer with a thickness of 400 nm was deposited by plasma-enhanced chemical vapor deposition (PECVD) at 350~$^\circ$C. The connection trenches were formed by Reactive Ion Etching (RIE), and 500 nm Al pad electrodes were fabricated by magnetron sputtering. Finally, individual detector chips were obtained by dicing the wafer using a diamond blade.

\subsection{Irradiation conditions}
Heavy-ion irradiation experiments were carried out at the Heavy Ion Research Facility in Lanzhou (HIFRFL) of the Institute of Modern Physics (IMP), Chinese Academy of Sciences. The detector was irradiated by 16 MeV/u $^{181}_{73}$Ta$^{35+}$ ions at room temperature under a vacuum of $10^{-5}$~Pa. The fluence and ion dose rate were $1\times10^{9}~\mathrm{cm}^{-2}$ and $1\times10^{7}~\mathrm{cm}^{-2}{s}^{-1}$, respectively. The total energy of Ta ion was 2.896 GeV.

\section{ELECTRICAL PERFORMANCE ANALYSIS}
 
The I-V characteristics of the devices were
measured using a Keithley 2470 source meter on a probe
station at room temperature. Fig. 2~(a) presents the I-V characteristics of the 4H-SiC PIN detector before and after $^{181}$Ta$^{35+}$ ion irradiation. Before irradiation, the leakage current of the detector is approximately $1.47 \times 10^{-10}$~A at a reverse bias voltage of 300 V. The underlying physical mechanism is that the wide bandgap of SiC ($\sim 3.2$~eV) significantly reduces the intrinsic carrier concentration, thereby effectively suppressing thermally activated leakage currents. After irradiation, the leakage current of the detector is approximately $1.49 \times 10^{-10}$~A at a reverse bias voltage of 300 V. The low leakage current of SiC detectors after heavy-ion irradiation results from the wide bandgap (~3.2~eV), which provides an ultralow intrinsic carrier concentration, and the deep-level nature of irradiation-induced defects, which exhibit low efficiency as generation-recombination centers. Additionally, the high atomic bonding energy of SiC ($\sim 4.6$~eV/atom) gives it a high displacement damage threshold, which inhibits the formation of leakage current-related defects induced by heavy ion irradiation. This demonstrates the excellent radiation hardness of the device, in sharp contrast to conventional silicon-based detectors, whose leakage currents typically increase dramatically under similar conditions due to enhanced bulk and surface leakage.

The $1/C^{2}\text{-}V$ characteristics were performed using a Keysight E4980A precision LCR meter with a  highvoltage bias adapter, with a test frequency of 10 kHz and AC signal amplitude of 50 mV. Fig. 2~(b) presents the $1/C^{2}\text{-}V$ characteristics of the 4H-SiC PIN detectors before and after heavy ion $^{181}_{73}$Ta$^{35+}$ irradiation. Before and after irradiation, the full depletion voltages of the 4H-SiC PIN detectors are consistently measured to be approximately 120 V. Before irradiation, the effective doping concentrations of the N-type epitaxial layer is approximately $6.23\times 10^{13}$~cm$^{-3}$, with a depletion depth of approximately 43~$\mu$m. The effective doping concentration has already approached the lowest level achievable by current SiC epitaxial growth technology, confirming that the doping and thickness design of the epitaxial layer fully meets the expected requirements. After irradiation, the effective doping concentrations of the N-type epitaxial layer is approximately $6.13\times 10^{13}$~cm$^{-3}$, with a depletion depth of approximately 44~$\mu$m. No significant changes in effective doping concentration, depletion voltage, or depletion depth are observed for the 4H-SiC PIN detector before and after $^{181}_{73}$Ta$^{35+}$ ion irradiation.This is mainly attributed to the synergistic effect of three factors. First, the low total fluence of heavy ion irradiation inherently limits the total number of displacement defects introduced via non-ionizing energy loss (NIEL)~\cite{Slurinskaya2002}. Second, the intense electronic excitation arising from the high linear energy transfer (LET) of heavy ions induces a dynamic annealing effect, prompting extensive in-situ recombination of point defects or their aggregation into electrically inactive clusters at room temperature, thus markedly lowering the net defect density. Third, the remaining defects are mostly deep-level acceptors (such as the carbon‑vacancy‑related Z$_{1/2}$ center), whose charge‑carrier trapping and compensation efficiency is far lower than that of shallow dopants, resulting in a negligible net impact on the effective doping concentration~\cite{Kawahara2014,Kawahara2013}.

Experimental results demonstrate that that for the unirradiated and $^{181}_{73}$Ta$^{35+}$ ion-irradiated 4H-SiC PIN detectors, the leakage current remains on the order of pA, while the depletion voltage and effective doping concentration remain stable. The indicates that the detector may offer advantages in low power consumption, radiation hardness, and high time resolution.

\begin{figure}[ht]
    \centering
    \includegraphics[width=1\linewidth]{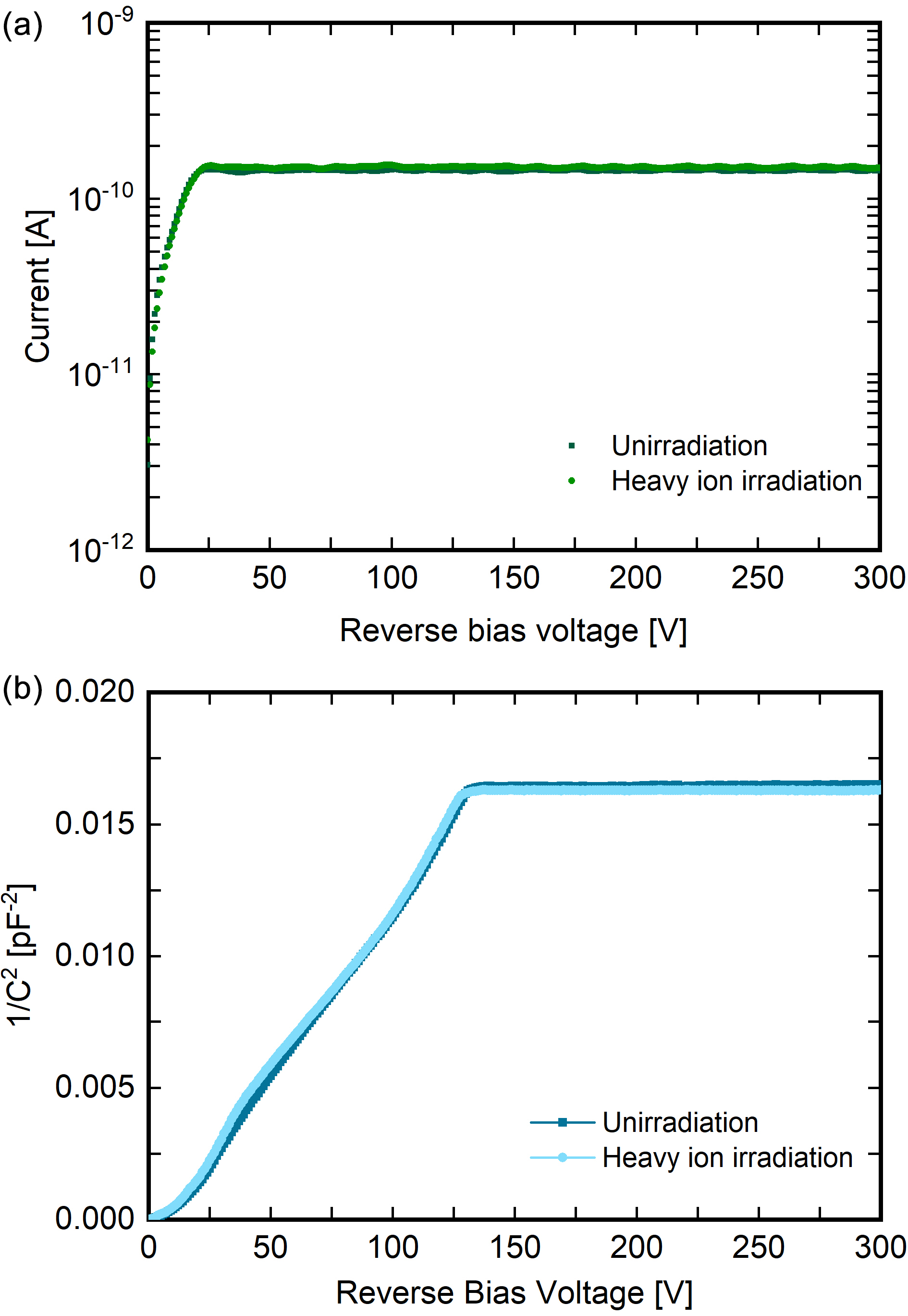}
    \caption{(a) I-V characteristics of 4H-SiC PIN detectors. (b) $1/C^{2}\text{-}V$ characteristics of 4H-SiC PIN detectors.}
    \label{fig:IV_CV}
\end{figure}

\section{Charge Collection Efficiency Performance}
\begin{figure}[ht]
    \centering
    \includegraphics[width=1\linewidth]{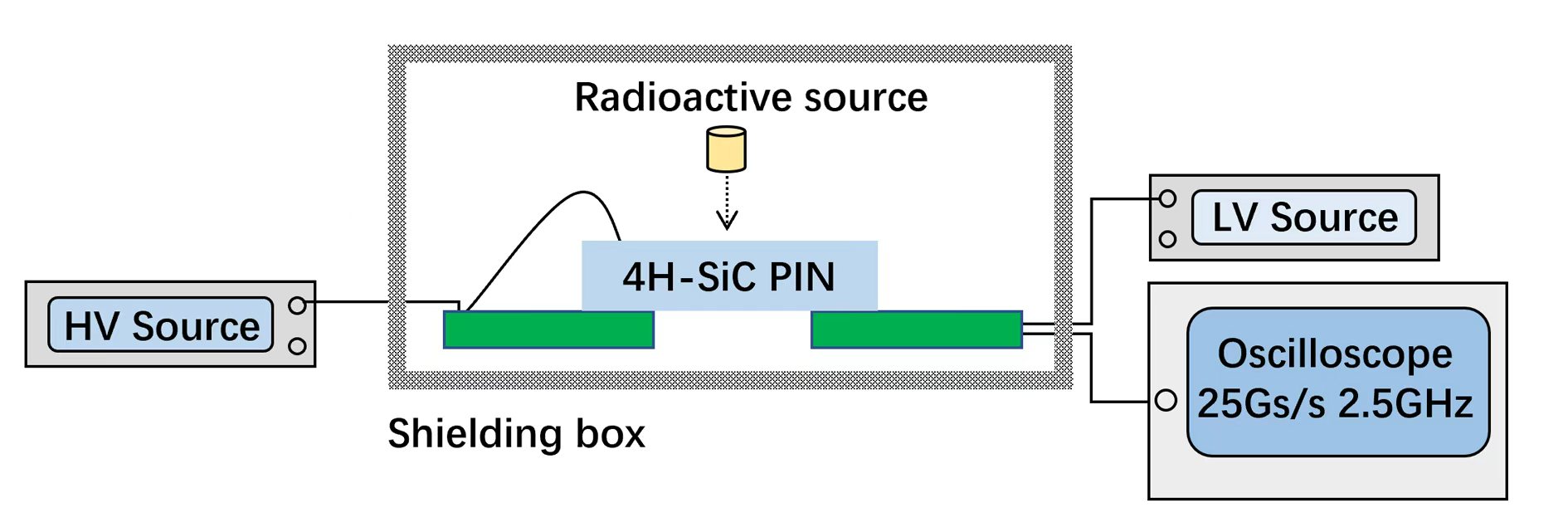}
    \caption{Experimental setup for charge collection efficiency of a $^{90}$Sr source.}
    \label{fig:sr_1}
\end{figure}
The charge collection performance setup for $\beta$ particles is shown
in Fig. 3. The experimental system consisted of a 
$^{90}$Sr radioactive source, a 4H-SiC PIN detector, a single-channel electronic readout board, a high-voltage source (Keithley 2470), a low-voltage source (GPD-3303, SGWINSTEK), and an oscilloscope (MSO64, Tektronix, 2.5 GHz). The detectors were encapsulated on the electronic readout board by conductive adhesive, and the pad electrodes of the detectors were electrically connected to the readout board. The high-voltage source supplied reverse bias to the detector, while the low-voltage source powered the single-channel electronic readout board.

\begin{figure}[ht]
    \centering
    \includegraphics[width=1\linewidth]{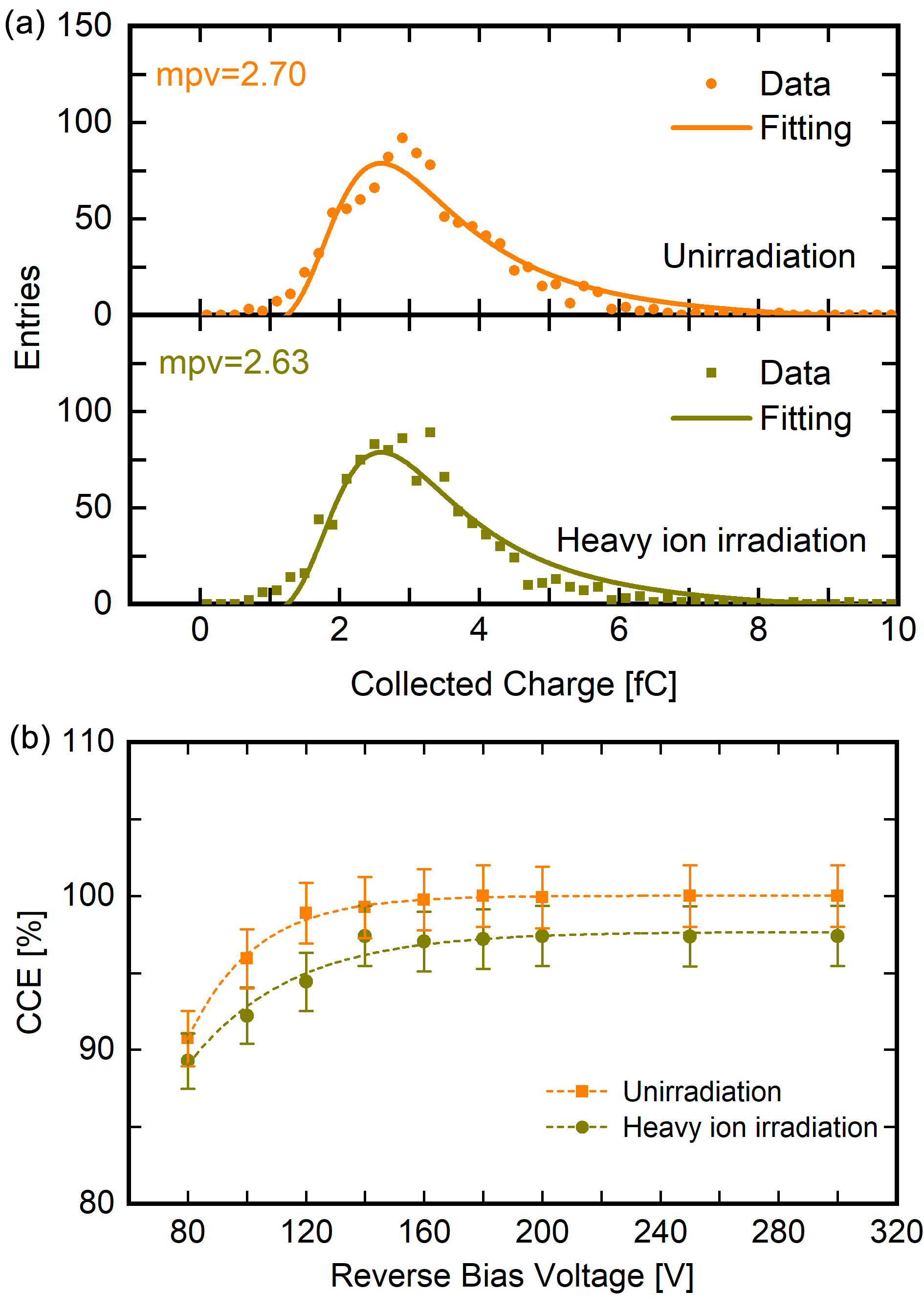}
    \caption{Charge collection performance of 4H-SiC PIN detectors at 300V: (a) Landau fit of the collected charge spectrum, with MPV indicating characteristic charge of detectors. (b) CCE versus reverse bias for the unirradiated and irradiated 4H-SiC PIN detectors}
    \label{fig:CCE}
\end{figure}

As shown in Fig.~4, the charge collection performances of the 4H-SiC PIN detector before and after ${}^{181}\mathrm{Ta}^{35+}$ ion irradiation were evaluated. Fig.~4(a) displays the charge collection spectra fitted with a Landau function. Due to the energy loss straggling caused beta particles traversing the sensitive region of the detector, the signal distribution exhibits a typical asymmetric Landau shape, which provides a more physically realistic description of energy deposition in thin semiconductor detectors compared to a symmetric Gaussian distribution. Under a reverse bias of 300 V, the most probable values (MPV) of the collected charge for the unirradiated detector and the ${}^{181}\mathrm{Ta}^{35+}$ ion irradiated detector are 2.70 fC and 2.63 fC, respectively. Fig. 4 (b) presents the CCE as a function of reverse bias voltage for these detectors. The full depletion voltages are about 120 V. For voltages greater than 120 V no further increase of the depletion depth occurs, and no increase in the collected charge is observed. The charge collection efficiency (CCE) of the unirradiated 4H-SiC PIN detector was defined as 100\% @ 300 V. The irradiated 4H-SiC PIN detector exhibited CCEs of 97.4\%@300~V. After irradiation, the 4H-SiC PIN detector achieved a CCE of 97.4\%@300~V, just 2.6\% lower than before irradiation. 

This minor degradation originates from the trapping and recombination of signal holes by deep-level carbon vacancy defects introduced by the low-fluence irradiation \cite{Klein2008,Hiyoshi2009}. The detector operates in hole-collection mode, where the signal mainly arises from the drift of holes toward the P$^{+}$ electrode. In the N$^{-}$ epilayer, the Z$_{1/2}$ deep acceptor associated with V$_{\rm C}$ is the dominant recombination center for minority holes. Its negatively charged state attracts positively charged holes via long-range Coulomb interaction, resulting in a large capture cross section. The hole mobility in 4H-SiC is on the order of only 100~cm$^{2}$V$^{-1}$s$^{-1}$. The slow drift velocity and the resulting long residence time in the depletion region further increase the trapping probability. Trapped holes undergo non-radiative recombination with electrons and are permanently lost, directly reducing the amount of collectable charge \cite{Kramberger2002}. The accompanying lattice disorder, such as Si–C bond rupture, introduces additional scattering centers that reduce hole mobility and prolong transit time, thereby indirectly increasing the trapping probability. After heavy‑ion irradiation, the SiC material exhibits a relatively low total defect concentration and no defect clusters have yet formed, leading to limited bulk recombination loss\cite{mao2024}. Consequently, the vast majority of signal holes in the SiC detector are still effectively collected by the electrode, and the charge collection efficiency remains at a high level\cite{Kramberger2002}. These results demonstrate that the 4H-SiC PIN detector maintains outstanding charge collection performance even after heavy ion irradiation.

\section{Time Resolution Performance}

\begin{figure}[ht]
    \centering
    \includegraphics[width=1\linewidth]{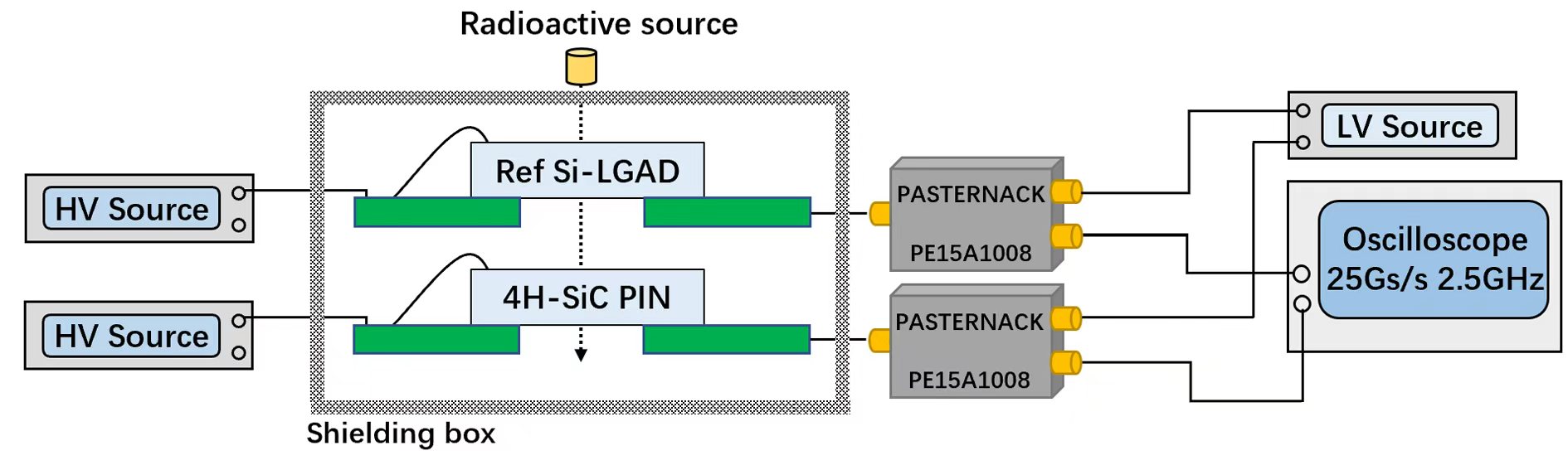}
    \caption{Experimental setup for time resolutin of a $^{90}$Sr source.}
    \label{fig:sr2}
\end{figure}

The time resolution experimental setup used for $\beta$ particles is illustrated in Fig.~5. This measurement system features a dual-channel configuration for time resolution measurements. The reference channel (REF) utilized a silicon low-gain avalanche detector (Si-LGAD, IHEP), which achieved a time resolution of 37 ps at an operating voltage of 200~V. The other channel served as the device under test (DUT) for mounting the 4H-SiC PIN detector whose time resolution was to be evaluated. Both detectors were concurrently exposed to $\beta$ particles emitted from the $^{90}$Sr source. Signals from both channels were read out using identical UCSC boards, which were then connected to PE15A1008 main amplifiers (20 dB gain). A hole of approximately 3 mm in diameter was drilled at the center of the bottom of the Si-LGAD readout board to allow $\beta$ particles to pass through. A high-voltage source was provided for each of the two channels to supply the reverse bias. When an incident particle passes through the detector, the resulting current signal is first processed by the pre-amplifier integrated on the readout board. A 20-dB main amplifier then provides further amplification of the signal. Signals from both channels were acquired using a high-speed oscilloscope featuring a single-channel sampling rate of 25 Gs/s. The generated synchronous pulse trigger signal was recorded by an oscilloscope, and the trigger jitter was found to be negligible.

The time difference between signals from the Device Under Test (DUT) and the reference (REF) is defined as

\begin{equation}
\Delta t = t_{\text{DUT}} - t_{\text{REF}}
\end{equation}

where $t_{\text{DUT}}$ and $t_{\text{REF}}$ represent the signal arrival times of the Si-LGAD and PIN detectors, respectively.

\begin{figure}[ht]
    \centering
    \includegraphics[width=1\linewidth]{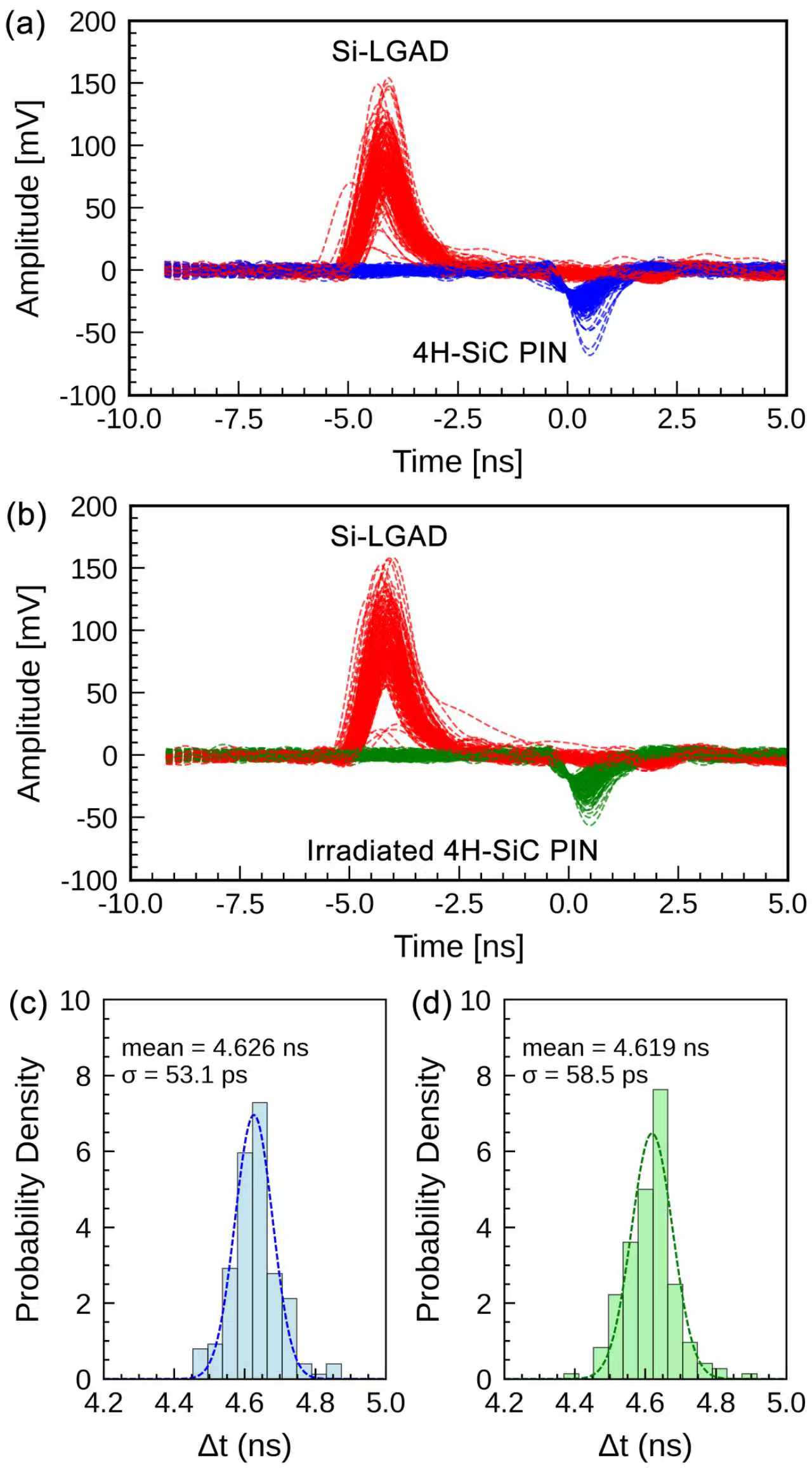}
    \caption{Reverse bias 300V: (a) Signal waveforms of Si LGAD and 4H-SiC PIN detectors. (b) Signal waveforms of Si LGAD and irradiated 4H-SiC PIN detectors.
   (c) Coincidence time resolution of 4H-SiC PIN detector. (d) Coincidence time resolution of irradiated 4H-SiC PIN detector.}
    \label{fig:wave_time}
\end{figure}

The time resolution is closely related to the probability distribution of the time difference t. To mitigate the influence of time walk variation in measured arrival time caused by differing signal amplitudes on the
time resolution, this paper employs the Constant Fraction Discrimination
(CFD) method for data processing. The CFD method determines
the trigger time at a constant fraction (50\%) of the signal amplitude, making the timing independent of pulse height. The process involves:

1) Finding the maximum amplitude Vmax and its time tmax.

2) Setting the threshold at Vth = 0.5 × Vmax.

3) Locating the point where the rising edge first crosses Vth and interpolating to obtain the precise CFD time.

The time spread $\sigma_{\text {$\Delta$t}}$ was extracted by performing a Gaussian fit to
the distribution of $\Delta$t. The timing resolution of the device under test, $\sigma_{DUT}$ was then calculated using:

\begin{equation}
\sigma_{DUT}=\sqrt{\sigma_{\Delta t}^{2}-\sigma_{Ref}^{2}}
\end{equation}

The timing spread, $\sigma_{\Delta t}$, was obtained by fitting a Gaussian function to the $\Delta t$ distribution. Subsequently, the timing resolution of the DUT was calculated using the expression:

\begin{equation}
\sigma_{\text{DUT}} = \sqrt{\sigma_{\Delta t}^{2} - \sigma_{\text{Ref}}^{2}}
\end{equation}

Fig.~6~(a) and Fig.~6~(b) show the signal waveforms of the Si LGAD and the unirradiated and irradiated 4H-SiC PIN detector at 300~V, respectively. 
Fig.~6~(c) and Fig.~6~(d) show the coincidence time resolutions of unirradiated and irradiated 4H-SiC PIN detectors. The coincidence time resolutions of the unirradiated and irradiated 4H-SiC PIN detectors are 53.1 ps and 58.5 ps, respectively. The time resolutions of unirradiated and irradiated 4H-SiC PIN detectors are 40~ps and 45~ps, from Equation (2). Its time resolution is better than that of state-of-the-art 4H-SiC low-gain avalanche detectors (LGADs) with a time resolution of 61 ps.

The total timing resolution $\sigma$ of the detector is given by the quadrature sum of its individual contributions:
\begin{equation}
\sigma^{2} = \sigma_{\rm TW}^{2} + \sigma_{\rm TDC}^{2} + \sigma_{\rm Jitter}^{2}
\label{eq4}
\end{equation}
where $\sigma_{\rm TW}$ is the time walk, which is eliminated in this study via the constant-fraction discrimination (CFD) method; $\sigma_{\rm TDC}$ arises from timing uncertainties in the analog-to-digital conversion, which is negligible at the 25~GS/s sampling rate employed.
The dominant term, $\sigma_{\rm Jitter}$, originates from noise in the detector and front-end electronics, and can be approximated as
\begin{equation}
\sigma_{\rm Jitter} \approx \frac{N}{dV/dt} \approx \frac{T_{\rm rise}}{S/N}
\label{eq5}
\end{equation}
where $N$ is the noise level, $dV/dt$ the signal slope at the discrimination threshold, $T_{\rm rise}$ the rise time, and $S/N$ the signal-to-noise ratio.
Equation~(\ref{eq5}) indicates that a larger signal amplitude and a shorter rise time both reduce the timing jitter, thereby enhancing the overall timing resolution.

\begin{figure}[ht]
    \centering
    \includegraphics[width=1\linewidth]{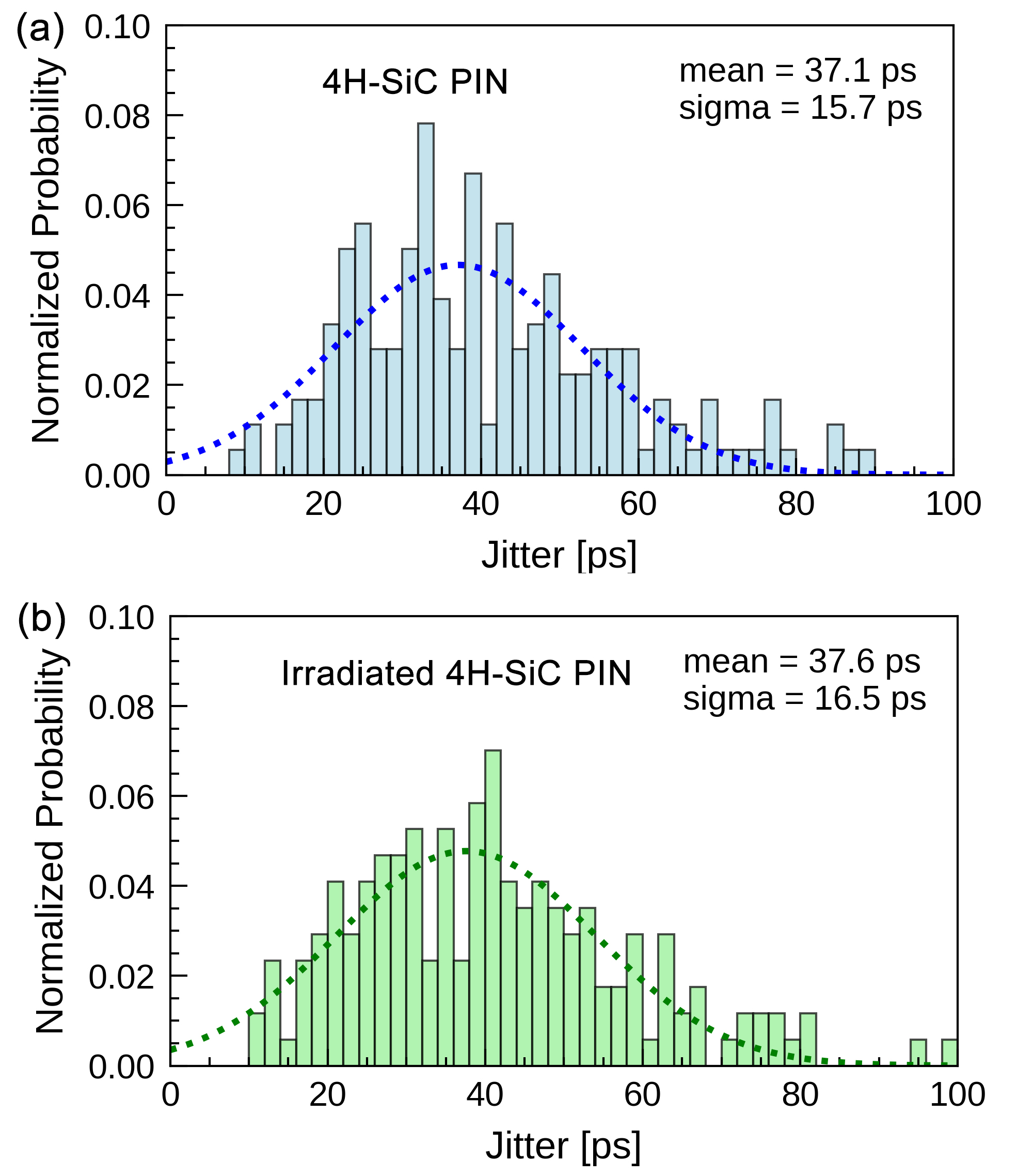}
    \caption{Jitter distribution: (a) 4H-SiC PIN detector. (b) Irradiated 4H-SiC PIN detector.}
    \label{fig:jitter}
\end{figure}

The minor change in time resolution may be attributed to the additional time fluctuation introduced by the random “capture-delayed release” process of signal holes by deep traps\cite{lutz2007,shockley1952}. Timing jitter, which dominates the time resolution, is primarily determined by the signal rise time and the S/N ratio. The Z$_{1}$/$_{2}$ traps produce two concurrent effects on holes: part of the trapped holes undergo recombination with electrons, leading to a reduction in signal amplitude S; the remaining trapped holes are slowly re-emitted via thermal excitation, giving rise to a slow decay tail on the pulse falling edge. This random "capture-release" process superimposes significant time dispersion onto the carrier transit time, directly increasing the jitter time\cite{lutz2007}. Meanwhile, the traps, acting as additional generation-recombination centers, marginally elevate the low-frequency noise. This marginal increase, along with the reduction in signal amplitude, jointly leads to further degradation of the S/N ratio\cite{bertuccio2002}.

\par Fig.~7 that the detector timing jitter  of unirradiated and irradiated 4H-SiC PIN detectors are 37.1~ps and 37.6~ps, respectively. However, the standard deviation of the jitter distribution ($\sigma \approx 16$~ps) remains essentially unchanged. This indicates that intrinsic transport parameters such as the saturation drift velocity of holes are not substantially affected by irradiation. The standard deviation of the jitter distribution exhibits little change, indicating that intrinsic transport parameters such as the saturation drift velocity of holes are not substantially affected. The degradation in time resolution is mainly due to the extra fluctuations caused by trap dynamics and the moderate decrease in S/N, not to a reduction in drift velocity. The experimental results show that the detector still retains a high time resolution of 45 ps irradiated with low fluence 16.5 MeV/u Ta Ions.

%This indicates that the intrinsic carrier transport parameters, such as the hole saturation drift velocity, are not significantly affected by irradiation. The degradation in timing resolution is primarily attributed to the additional fluctuation introduced by trap dynamics and the limited reduction in signal-to-noise ratio, rather than a decrease in the drift velocity itself. After irradiation, the detector still maintains a sub-45~ps timing resolution.%

\section{Conclusion}

A silicon carbide PIN detector was fabricated and its radiation tolerance under Ta heavy ion irradiation of 2370 MeV was evaluated. Its electrical properties, charge collection performance and time resolution of  $\beta$-particles ($^{90}$Sr) are reported. The leakage currents for unirradiated and irradiated 4H-SiC PIN detectors are $1.47 \times 10^{-10}$~A @ 300 V and 1.49~$\times$ 10$^{-10}$A@ 300 V.
The effective doping concentrations for unirradiated and irradiated 4H-SiC PIN detectors are $6.23\times 10^{13}$~cm$^{-3}$ and $6.13\times 10^{13}$~cm$^{-3}$. The irradiated detector exhibits good electrical performance and stable device architecture. The 4H-SiC PIN detector exhibits a charge collection efficiency (CCE) of 99.24\% under Ta Heavy Ion Irradiation. The time resolutions of the detector before and after irradiation are 40 ps and 45 ps, respectively. Experimental results indicate that the CCE and time resolution performance exhibit good stability before and after irradiation. These results demonstrate stable performance under extreme Ta Heavy Ion Irradiation, highlighting the detectors potential for heavy ion radiation‑hard applications in high-energy physics, space missions, and nuclear reactor monitoring.

Our team has successfully fabricated graphene-optimized SiC PiN and SiC LGAD detectors. In the near future, we will also research on the effects of carbon ions on the charge collection efficiency and time resolution of 4H-SiC PIN and 4H-SiC LGAD detectors. In addition, we will continue to advance the research and development of SiC-BJT detector. The SiC BJT detector explores potential applications in detector readout electronics and signal amplification. The development of these novel devices will further enrich the SiC detector technology portfolio and promote their engineering applications in four-dimensional tracking, time-resolved imaging, and extreme radiation environments.

\section{Acknowledgement}
We acknowledge the RASER team (https://raser.team) and CERN DRD3 Collaboration (https://drd3.web.cern.ch) for their useful discussions.

\bibliographystyle{ieeetran}
\bibliography{references}

\end{document}